\documentclass[12pt,a4paper,final]{iopart}

\usepackage{iopams}
\expandafter\let\csname equation*\endcsname\relax
\expandafter\let\csname endequation*\endcsname\relax
\usepackage{amsmath,bm}
\usepackage{graphicx}
\usepackage{caption}
\usepackage{subcaption}
\usepackage{MnSymbol}
\usepackage[breaklinks=true,colorlinks=true,linkcolor=blue,urlcolor=blue,citecolor=blue]{hyperref}

\usepackage{epstopdf}
\usepackage{dcolumn}
\usepackage{bm}
\usepackage{setspace}
\usepackage{psfrag}
\usepackage{float} 
\newcommand{\beq}{\begin{eqnarray}}
\newcommand{\eeq}{\end{eqnarray}}

\usepackage{geometry}
\usepackage{amssymb}
\usepackage{esint}
\usepackage[english]{babel}
\usepackage[T1]{fontenc}
\usepackage[latin9]{inputenc}

\newcommand{\ThreeJ}[6]{\ensuremath{\begin{pmatrix}#1 & #2 & #3\\ #4 & #5 & #6\end{pmatrix}}}

\newcommand{\WignerD}[3]{\ensuremath{\mathcal{D}^{#1}_{#2 #3}}}

\begin{document}

\title[General theory of PXECD]{General theory of photoexcitation induced photoelectron circular dichroism}

\author{Alex G Harvey$^{1}$, Zden\v{e}k Ma\v{s}\'{i}n$^{1}$ and  Olga Smirnova$^{1,2}$}
\address{$^1$Max-Born-Institut, Max-Born-Str. 2A, 12489 Berlin}
\address{$^2$ Technische Universit{\"a}t Berlin,
Ernst-Ruska-Geb{\"a}ude, Hardenbergstr. 36A,10623, Berlin, Germany.}

\begin{abstract}
The photoionization of chiral molecules prepared in a coherent superposition of excited states 
can  give access to the underlying chiral coherent dynamics in a procedure known as photoexcitation induced photoelectron circular dichroism (PXECD) \cite{PXECDp1,beaulieu2016,beaulieu2018}. 
This exclusive dependence on coherence can also be seen in a different part of the angular spectrum, where it is not contingent on the chirality of the molecule, thus allowing extension of PXECD's sensitivity to tracking coherence to non-chiral molecules.
Here we present a general theory of PXECD based on angular momentum algebra and derive  explicit expressions for all pertinent asymmetry parameters which arise  for arbitrary polarisation of pump and probe pulses. 
The theory is  developed in a way that clearly and simply separates chiral and non-chiral contributions to the photoelectron angular distribution, and also demonstrates how PXECD and PECD-type contributions, which may be distinguished by whether pump or ionizing probe enables chiral response,
are mixed when arbitrary polarization is used.
\end{abstract}

\begin{figure}
\global\long\def\del{\boldsymbol{\nabla}}

\global\long\def\x{\times}

\global\long\def\t{\cdot}

\global\long\def\d{\mathrm{d}}

\global\long\def\ket#1{\left|#1\right\rangle }

\global\long\def\bra#1{\left\langle #1\right|}

\global\long\def\braket#1#2{\langle#1|#2\rangle}

\global\long\def\vecdiptens#1#2{\mathbf{D}_{#1}^{\mathrm{M}#2} }

\global\long\def\elmdiptens#1#2{D_{#1}^{\mathrm{M}#2} }

\global\long\def\vecbound#1#2{\mathbf{d}_{#1}^{\mathrm{M}#2} }

\global\long\def\elmbound#1#2{d_{#1}^{\mathrm{M}#2} }

\global\long\def\vecbigA{\mathbf{A}^{\mathrm{M}} }

\global\long\def\alphazero#1{D^{\mathrm{M}}_{ii',(#1)00}(E) \mathbf{d}^{M}_{i0} \cdot \mathbf{d}^{M}_{i'0}}
\global\long\def\alphaone#1{\mathbf{D}^{\mathrm{M}\dagger}_{ii',(#1)1}(E) \cdot (\mathbf{d}^{M}_{i0} \times \mathbf{d}^{M}_{i'0}) }
\global\long\def\alphatwo#1{\mathbf{D}^{\mathrm{M}\dagger}_{ii',(#1)2}(E) \cdot \mathbf{A}_{ii',2}}

\global\long\def\da{\dagger}

\global\long\def\avg#1{\left\langle #1\right\rangle }

\global\long\def\h{\hbar}

\global\long\def\eval#1{\left.#1\right|}

\global\long\def\b#1{\boldsymbol{#1}}

\global\long\def\txt#1{\mathrm{#1}}

\global\long\def\i{\mathrm{i}}

\global\long\def\e{\mathrm{e}}

\global\long\def\ra{\rightarrow}
\end{figure}
\section{Introduction}
Chirality is associated with mirror-symmetry breaking. It is ubiquitous in nature and  fundamental to the understanding of natural processes.
For chiral molecules, this mirror-symmetry breaking leads to two versions of a molecule, the left and right enantiomers.

Today, characterising molecular chirality is a dynamic and multidisciplinary research field with an expanding arsenal of techniques. In the gas phase, these include techniques such as Coulomb explosion imaging \cite{pitzer2013}, microwave detection \cite{patterson2013}, the combination of mass spectrometry with multiphoton and vibrational excitation techniques \cite{lux2012,rhee2009}, high harmonic spectroscopy \cite{cireasa2015,smirnova_opportunities_2015,ayuso2018}, and photoelectron circular dichroism (PECD) \cite{ritchie1976,powis2000a,powis2000a,garcia2013}.
The growing interest in the response of chiral molecules in the time domain has motivated recent and ongoing efforts to develop non-linear chiroptical techniques  in optical \cite{fischer2005,abramavicius2006,choi_two-dimensional_2008,fidler2014} and XUV domain \cite{rouxel2017photoinduced} to that aim.  

PECD relies on the difference in angular resolved photo-electron emission for left and right circlarly polarized light. 
Due to the extra directionality coming from observation of the photoelectron, dichroism can be seen, after orientational averaging, in the electric dipole approximation; consequently the strength of the dichroism is significantly greater, on the order of $10$\% of the total signal \cite{powis2000}, than techniques reliant on magnetic dipole effects.
In both single and multi-photon PECD, the highest PECD signal is seen in the low-energy region of the spectrum.
PECD shows a strong dependence on both the initial, intermediate and final states and is a structurally sensitive probe, as seen in the striking difference observed between camphor and fenchone due to the methyl group substitution, although the involved bound states and photoelectron spectra hardly change \cite{powis2008}, and in the pronounced dependence of PECD signal on molecular geometry and sensitivity to non-Frank-Condon effects \cite{garcia2013}. 

In contrast to conventional PECD, PXECD requires the coherent population of multiple states, and hence the dichroic signal displays quantum beating with respect to the delay between excitation and ionization pulses  \cite{beaulieu2016,beaulieu2018}. 
PXECD is thus a form of time-resolved photoelectron spectroscopy (TRPES) and can be used to investigate the time evolution of various intramolecular processes (for a review see \cite{stolow_time-resolved_2008-1}). 
TRPES has its origins in studies in which atomic hyperfine levels were coherently excited and probed by ionization at nanosecond delay using linear pulses \cite{strand_influence_1978,leuchs_quantum_1979,chien_angular_1983}. 
This lead to the observation of quantum beats in the photoelectron angular distributions and allowed information on the ionization continuum and the hyperfine interaction to be extracted. 
Later work extended this concept to the hyperfine levels of the NO molecule \cite{reid_observation_1994}. 
As shorter pulses became available, experimental and numerical studies involving the coherent excitation of rotational states in the first step examined the influence of rotation-vibration coupling \cite{reid_photoelectron_1999,althorpe_predictions_2000}, and non-adiabatic dynamics \cite{underwood_time-resolved_2000} in small molecules at pico to femtosecond time resolution.
Recent TRPES studies include joint experimental and theoretical work to time-resolve valence electron dynamics during a chemical reaction \cite{hockett_time-resolved_2011}, and a theoretical study of non-adiabatic dynamics in the vicinity of a conical intersection \cite{bennett_nonadiabatic_2016}.
We anticipate PXECD to be a similarly useful tool with the added bonus of sensitivity to the chirality of the studied system.

In this paper we extend and generalise our previous theoretical descriptions of PXECD,  combining the best aspects of our initial angular algebra based approach \cite{PXECDp1} and our later approach in \cite{beaulieu2016,beaulieu2018}, and offer a complementary perspective on this phenomenon.

\section{Theory}
As in our previous works \cite{PXECDp1,beaulieu2016,beaulieu2018}, we model the interaction
between the electric field and the molecule using first
order perturbation theory and the dipole approximation. 

We define the pump field in the laboratory reference frame as:
\begin{equation}
\mathbf{E}^{\mathrm{L}}\left(t\right)=\frac{1}{\sqrt{2}}F\left(t\right)\hat{\bm{\varepsilon}}^{\mathrm{L}}\e^{-\i\left(\omega t+\delta\right)}+\mathrm{c.c.} 
\end{equation}
where $\omega$ is the carrier frequency, $F(t)$ includes the field amplitude and the envelope, 
and the carrier-envelope phase $\delta$ determines the orientation of the electric field
vector at the moment $t=0$. Finally,  the helicity  $\sigma=\pm 1$ determines whether the field is left or right polarized and the polarization of the field is expressed in the spherical basis
\begin{equation}
\hat{\varepsilon}_{-1}^{\mathrm{L}}=\frac{1}{\sqrt{2}}(\hat{x}^{\mathrm{L}}-\i\hat{y}^{\mathrm{L}}),
\qquad \hat{\varepsilon}_{0}^{\mathrm{L}}= \hat{z}^{\mathrm{L}} \qquad \hat{\varepsilon}_{+1}^{\mathrm{L}}=\frac{-1}{\sqrt{2}}(\hat{x}^{\mathrm{L}}+\i\hat{y}^{\mathrm{L}})
\end{equation}
The superscripts $\mathrm{L}$ and  $\mathrm{M}$ indicate vectors are in the laboratory and molecular frame respectively. 
The transformation
of vectors from the lab frame to the molecular frame is performed
according to $\mathbf{v}^{\mathrm{M}}=\mathbf{D}^{\dagger}\left(\varrho\right)\mathbf{v}^{\mathrm{L}}$ where $\varrho\equiv\left(\alpha,\beta,\gamma\right)$ the Euler
angles in the active $z$-$y$-$z$ convention.
In the angular momentum basis, this rotation operator corresponds to the Wigner rotation matrix, we use its complex transpose here to account for the usual convention that the Wigner rotation matrix transforms basis vectors covariantly.
Using perturbation theory, after the end of the pump pulse of duration $T_{1}$, 
we find the wave function at a time $\tau$:
\begin{equation}
\psi_{\varrho}\left(\tau\right)=c_{0}\psi_{0}\e^{-\i\omega_{0}\tau}+\sum_{i=1}c_{i}\left(\varrho\right)\psi_{1}\e^{-\i\omega_{i}\tau},
\end{equation}
where $c_{0}\approx1$ and the expressions for the excitation amplitudes are standard: 
\begin{equation}
c_{i}\left(\varrho\right)=\i\left[\mathbf{d}_{i0}^\mathrm{M}\t \mathbf{D}^\dagger\left(\varrho\right)\hat{\bm{\varepsilon}}^{\mathrm{L}}\right]\mathcal{E}\left(\omega_{i0}\right)
\end{equation}
$i$ labels the intermediate excited states, $\mathbf{d}_{i0}^{\mathrm{M}}$ are the transition dipole matrix elements to these states from the ground state, in the molecular frame spherical basis.
The excitation amplitude is proportional to the spectral component of the pump at the corresponding transition frequency $\omega_{i0}$, $\mathcal{E}\left(\omega_{i0}\right)$. 

To calculate the photoelectron angular distribution resulting from the photoionization of excited states we need to consider the bound-free transitions due to the probe field with polarization $\hat{\bm{\xi}}^\mathrm{L}$.
The population amplitude of a continuum state $\mathbf{k}^\mathrm{M}$ after the end of the 
probe pulse, assuming that the pump and the probe do not overlap, is
\begin{eqnarray}
c(\mathbf{k}^\mathrm{M};\varrho)&=&\i \sum_ic_{i}(\varrho)\e^{-\i\omega_{i}\tau}\left[\mathbf{d}_{i}^\mathrm{M}(\mathbf{k}^\mathrm{M})\t \mathbf{D}^\dagger\left(\varrho\right)\hat{\bm{\xi}}^{\mathrm{L}}\right]\mathcal{E}^{\prime}\left(\omega^{\prime}_{\mathbf{k}i}\right)
\label{ProbeAmpl}
\end{eqnarray}
where $\mathcal{E}^{\prime}\left(\omega_{\mathbf{k},i}\right)$ is the spectral amplitude of the probe at the 
required transition frequency
and $\mathbf{d}_{i}^\mathrm{M}(\mathbf{k})$ are bound-free transition dipoles in the molecular frame.
In this work we will consider electronic states only.
The molecular frame PAD is proportional to 
\begin{eqnarray}
\frac{d\sigma}{d\mathbf{k}^\mathrm{M}}(\varrho,\tau)&\propto& \left|  \sum_i\e^{-\i\omega_{i}\tau}\left[\mathbf{d}_{i}^\mathrm{M}(\mathbf{k}^\mathrm{M})\t\mathbf{D}^\dagger\left(\varrho\right)\hat{\bm{\xi}}^{\mathrm{L}}\right]\left[\mathbf{d}_{i0}^\mathrm{M}\t\mathbf{D}^\dagger\left(\varrho\right)\hat{\bm{\varepsilon}}^{\mathrm{L}}\right]  \right|^2
\label{ProbeAmpl2}
\end{eqnarray}
Performing a partial wave expansion for the photoelectron and writing component-wise
\begin{eqnarray}
\frac{d\sigma}{d\mathbf{\hat k}^\mathrm{M}}(E,\tau;\varrho) &\propto& \left| \sum_i\e^{-\i\omega_{i}\tau}\sum_{lmp_2q_2}\WignerD{1*}{p_2}{q_2}\hat{\xi}^\mathrm{L}_{p_2}d^{\mathrm{M}}_{i,q_2,lm}(E)Y_{lm}(\mathbf{\hat k}^M)\sum_{p_1q_1}\WignerD{1*}{p_1}{q_1}\hat{\varepsilon}^\mathrm{L*}_{p_1} d^{M}_{i0,q_1} \right|^2,
\end{eqnarray}
We note that we have absorbed a factor of $i^{-l}e^{i\sigma_l}$, where $\sigma_l$ is the Coulomb phase, into the dipole matrix elements in contrast to how they are usually written. Expanding the modulus square we get, 
\begin{align}
\frac{d\sigma}{d\mathbf{\hat k}^\mathrm{M}}(E,\tau;\varrho) \propto  \sum_{ii'}\e^{-\i\omega_{ii'}\tau}&\sum_{K_eM_e}\sum_{lmp_2q_2}\WignerD{1*}{p_2}{q_2}d^{\mathrm{M}}_{i,q_2,lm}(E)d^{\mathrm{M}*}_{i',q'_2,l'm'}(E)\WignerD{1}{p'_2}{q'_2}\rho^{\xi\mathrm{L}}_{p_2p'_2} Y_{K_eM_e}(\mathbf{\hat k}^M) \nonumber \\ 
&\times (-1)^{m'+M_e}\left[\frac{\tilde l\tilde l'\tilde K_e}{4\pi}\right]^{1/2}\ThreeJ{l}{l'}{K_e}{-m}{m'}{M_e}\ThreeJ{l}{l'}{K_e}{0}{0}{0} \nonumber \\
&\sum_{p_1q_1}\WignerD{1*}{p_1}{q_1} d^{M}_{i0,q_1}d^{M*}_{i'0,q'_1}\WignerD{1}{p'_1}{q'_1}\rho^{\varepsilon\mathrm{L}}_{p_1p'_1},
\end{align}
where the product of polarization vectors $\hat{\varepsilon}^{\mathrm{L}}_{p_1}\hat{\varepsilon}^{\mathrm{L}*}_{p'_1}=\rho^{\varepsilon\mathrm{L}}_{p_1p'_1}$ and $\hat{\xi}^{\mathrm{L}}_{p_2}\hat{\xi}^{\mathrm{L}*}_{p'_2}=\rho^{\xi\mathrm{L}}_{p_2p'_2}$ give elements of the polarization density matrix for the first and second photon, and the product of spherical harmonics has been contracted using the identity
\begin{eqnarray}\label{SpHContraction}
&&Y_{lm}(\mathbf{\hat k})Y^{*}_{l'm'}(\mathbf{\hat k})= \sum_{K_eM_e}(-1)^{m}\left[\frac{\tilde l\tilde l'\tilde K_e}{4\pi}\right]^{1/2}\ThreeJ{l}{l'}{K_e}{-m}{m'}{M_e}\ThreeJ{l}{l'}{K_e}{0}{0}{0} Y_{K_e M_e}(\mathbf{\hat k}).\nonumber\\
\end{eqnarray}
involving the $3-j$ symbols and where $\tilde l=2l+1$. At this point it is useful to introduce some of the properties of the $3-j$ symbol, as they will be crucial later. They have a simple relation to the Clebsch-Gordon coefficients used to couple angular momentum (see, for example \cite{brinksatchler}), but treat each angular momentum vector on an equal footing, instead of coupling two angular momenta to give a third, they couple three angular momenta to give a scalar invariant, $\sum_{abc}\ket{Aa}\ket{Bb}\ket{Cc}\ThreeJ{A}{B}{C}{a}{b}{c}=\ket{00}$. An important symmetry property is that a $3-j$ symbol is unchanged after even permutation of its column, and acquires a phase $(-1)^{(A+B+C)}$ under odd permutations, the same phase is acquired if the bottom row is multiplied by $-1$ (equivalent to inversion in 3D). From this it can be seen that if the sum of the top row is odd (and the three vectors are polar) then the scalar invariant is a pseudo-scalar. This is the hall mark of a chiral quantity and we will now proceed to transform the equation for the PAD into a form in which this can be seen explicitly. 

With this in mind we observe that the product of dipoles and the product of the polarization density matrix and spherical harmonic are themselves elements of tensors that can be put in spherical tensor form. The general form of this transformation is $d_{Cc}=\sum_{ab}(-1)^a\tilde C^{\tfrac{1}{2}}\ThreeJ{A}{B}{C}{-a}{b}{c}d_{Aa,Bb}$ We also rotate the outgoing electron direction into the lab frame where it is detected.
\begin{align}\label{eqn_dxs_before_avg}
\frac{d\sigma}{d\mathbf{\hat k}^\mathrm{L}}(E,\tau;\varrho) \propto  \sum_{ii'}\e^{-\i\omega_{ii'}\tau}&\sum_{\substack{K_2K_eK_J\\M_JN_J}}D^{\mathrm{M}}_{ii',(K_2K_e)K_JM_J}(E)\WignerD{K_J*}{N_J}{M_J} Z^{(K_2K_e)}_{K_JN_J}(\mathbf{\hat k}^L) \nonumber \\ 
&\sum_{\substack{p_1q_1\\p'_1q'_1}}\WignerD{1*}{p_1}{q_1} d^{M}_{i0,q_1}d^{M*}_{i'0,q'_1}\WignerD{1}{p'_1}{q'_1}\rho^{\varepsilon\mathrm{L}}_{p_1p'_1},
\end{align}
where, 
\begin{align}\label{eqn_def_D}
D^{\mathrm{M}}_{ii',(K_2K_e)K_JM_J}(E)=\sum_{M_eM_2}(-1)^{M_2}\tilde K_J^{\tfrac{1}{2}}\ThreeJ{K_2}{K_e}{K_J}{-M_2}{M_e}{M_J} \sum_{q_2q_2'}(-1)^{q_2}\tilde K_2^{\tfrac{1}{2}}\ThreeJ{1}{1}{K_2}{-q_2}{q_2'}{M_2}\nonumber\\
\sum_{\substack{ll'\\mm'}}(-1)^{m}\tilde K_e^{\tfrac{1}{2}} \ThreeJ{l}{l'}{K_e}{-m}{m'}{M_e}d^{\mathrm{M}}_{i,q_2,lm}(E)d^{\mathrm{M}*}_{i',q'_2,l'm'}(E) \ThreeJ{l}{l'}{K_e}{0}{0}{0}\left(\frac{ \tilde l \tilde l' }{4\pi}\right)^{\tfrac{1}{2}}
\end{align}
and
\begin{align}
Z^{(K_2K_e)}_{K_JN_J}(\mathbf{\hat k}^L)=\sum_{\substack{N_eN_2\\p_2p'_2}}(-1)^{N_2+p_2}(\tilde K_2 \tilde K_J)^{\tfrac{1}{2}} \ThreeJ{K_2}{K_e}{K_J}{-N_2}{N_e}{N_J}\ThreeJ{1}{1}{K_2}{-p_2}{p'_2}{N_2}\rho^{\xi\mathrm{L}}_{p_2p'_2} Y_{K_eN_e}(\mathbf{\hat k}^{\mathrm{L}}) 
\end{align}
To get the PAD from a randomly oriented gas sample, we must orientationally average eqn. \ref{eqn_dxs_before_avg}:
\begin{eqnarray}
\frac{\overline{d\sigma}}{d\mathbf{\hat k}^\mathrm{L}}(E,\tau) = \frac{1}{8\pi^2}\int \frac{d\sigma}{d\mathbf{\hat k}^\mathrm{L}}(E,\tau;\varrho) d\varrho
\end{eqnarray}
giving
\begin{align}
\frac{\overline{d\sigma}}{d\mathbf{\hat k}^\mathrm{L}}(E,\tau) \propto  \sum_{ii'}\e^{-\i\omega_{ii'}\tau}&\sum_{\substack{K_2K_eK_J\\N_Jp_1p'_1}} \left\{\sum_{M_Jq_1q'_1}D^{\mathrm{M}}_{ii',(K_2K_e)K_JM_J}(E) d^{M}_{i0,q_1}d^{M}_{i'0,-q'_1}\ThreeJ{K_J}{1}{1}{M_J}{q_1}{-q_1'}\right\} \nonumber \\ 
& (-1)^{p'_1}\ThreeJ{K_J}{1}{1}{N_J}{p_1}{-p_1'}\rho^{\varepsilon\mathrm{L}}_{p_1p'_1}Z^{(K_2K_e)}_{K_JN_J}(\mathbf{\hat k}^L),
\end{align}
The PAD has been separated into two parts: outside the braces are the lab frame quantities (the photon polarizations and the outgoing electron direction), inside the braces we get a scalar invariant involving the molecular frame quantities only, namely the transition and ionization dipoles. We denote this invariant scalar $\alpha^{(K_2K_e)K_J}$ where $K_J$ can take the values $\{0,1,2\}$, and also transform the photon density matrices into their irreducible spherical tensor form.
The non-vanishing components of the first photon density matrix are: 
$\rho^{\varepsilon\mathrm{L}}_{00}=-\sqrt{1/3}$, $\rho^{\varepsilon\mathrm{L}}_{10}=-\sqrt{1/2}C_1$, $\rho^{\varepsilon\mathrm{L}}_{20}=-\sqrt{1/6}$ and $\rho^{\varepsilon\mathrm{L}}_{22}=\rho^{\varepsilon\mathrm{L}*}_{2-2}=(1/2)L_1$. Here $-1\le C_1 \le 1$ defines the the amount of circular polarization and $0\le L_1 \le 1$ the amount of linear polarization. $L^2_1+C^2_1$ is unity for pure polarization, less than 1 for partial polarization and 0 for unpolarized (in the $x-y$ plane) light. The major axis of polarization defines the $x$-direction and the propagation direction is $z$.
\begin{align}\label{eqn:simplest_general}
\frac{\overline{d\sigma}}{d\mathbf{\hat k}^\mathrm{L}}(E,\tau) \propto  \sum_{ii'}\e^{-\i\omega_{ii'}\tau}&\sum_{\substack{K_2K_eK_J\\N_J}} \alpha^{(K_2K_e)K_J} \rho^{\varepsilon\mathrm{L}}_{K_JN_J}Z^{(K_2K_e)}_{K_JN_J}(\mathbf{\hat k}^L),
\end{align}
We can write this in vector form as 
\begin{align}
\alpha^{(K_2K_e)0}&=\mathbf{D}^{\mathrm{M}\dagger}_{ii',(K_2K_e)0}(E) \cdot \mathbf{A}_{ii',0}&&=-\tfrac{1}{\sqrt{3}}D^{\mathrm{M}}_{ii',(K_2K_e)00}(E) \mathbf{d}^{M}_{i0} \cdot \mathbf{d}^{M}_{i'0} \nonumber \\
\alpha^{(K_2K_e)1}&=\tfrac{1}{\sqrt{3}}\mathbf{D}^{\mathrm{M}\dagger}_{ii',(K_2K_e)1}(E) \cdot \mathbf{A}_{ii',1}&&=\tfrac{1}{\sqrt{6}} \mathbf{D}^{\mathrm{M}\dagger}_{ii',(K_2K_e)1}(E) \cdot (\mathbf{d}^{M}_{i0} \times \mathbf{d}^{M}_{i'0}) \nonumber \\
\alpha^{(K_2K_e)2}&=\tfrac{1}{\sqrt{5}}\mathbf{D}^{\mathrm{M}\dagger}_{ii',(K_2K_e)2}(E) \cdot \mathbf{A}_{ii',2},&
\end{align}
where
\begin{align}
A_{ii',K_JM_J}=\sum_{q_1q'_1} (-1)^{q_1'}\tilde K_J^{\tfrac{1}{2}}d^{M}_{i0,q_1}d^{M*}_{i'0,q'_1}\ThreeJ{K_J}{1}{1}{M_J}{q_1}{-q_1'}.
\end{align}
We can now give a physical interpretation to the various irreducible spherical tensors  above. 
In general, expression of the quantities above in terms of irreducible spherical tensors is a multipole expansion \cite{fano}. 
The  zeroth order tensor is a scalar and thus isotropic, the first order tensor is known as the orientation vector, it corresponds to an net orientation of the angular momentum of the system, and has the form of a dipole, the second order tensor is known as the alignment vector and is of quadrupole form (see, for example \cite{blum}).

We see that $ \mathbf{A}_{ii',K_J}$ describes isotropy/orientation/alignment of the system induced by the first pulse, while $\mathbf{D}^{\mathrm{M}\dagger}_{ii',(K_2K_e)K_J}(E)$ describes the further orientation/alignment induced by the second pulse and detection of the photoelectron.
PXECD arises from the $\alpha^{(K_2K_e)1}$ coefficient, therefore we see that orientation of the ensemble by the pump pulse is integral to PXECD.
This orientation creates an induced net dipole in the ensemble that oscillates with angular frequency $\omega_{ii'}$ as discussed in \cite{beaulieu2016,beaulieu2018}.

It is easy to see that $\alpha^{(K_2K_e)1}$ exist only when the bound transition dipoles are non-parallel and hence only exists for the interference terms (involving different excited states) not the direct terms.
This implies it requires coherent population of multiple states to be observed.
One might be tempted to say that $\alpha^{(K_2K_e)K_J}$ is scalar for even values of $K_J$ and pseudoscalar for odd values, but some care must be taken here, the $\mathbf{d}^{M}_{i0}$ are polar vectors, however $\mathbf{D}^{\mathrm{M}}_{ii',(K_2K_e)K_J}(E)$ can be either a polar or pseudovector, examination of the $3$j symbols in eqn. \ref{eqn_def_D} shows that under inversion $\mathbf{i}\mathbf{D}^{\mathrm{M}}_{ii',(K_2K_e)K_J}(E) = (-1)^{K_e+K_J}\mathbf{D}^{\mathrm{M}}_{ii',(K_2K_e)K_J}(E)$ i.e. it is a pseudovector when $K_e+K_J$ is odd. 
Hence  $\alpha^{(K_2K_e)K_J}$ is a pseudoscalar only when $K_e$ is odd (note: the dot product of a vector and a pseudovector is a pseudoscalar while the dot product of a pseudovector and a pseudovector is a scalar).

It is straightforward to demonstrate that pseudoscalar $\alpha^{(K_2K_e)K_J}$ exist only in chiral molecules. Consider first the reflection of a randomly oriented ensemble of chiral molecules. This operation changes the sign of pseudoscalar $\alpha^{(K_2K_e)K_J}$ but also changes the enantiomer in the sample. 
Pseudoscalar $\alpha^{(K_2K_e)K_J}$ is therefore the source of asymmetry in photoelectron emission that changes sign with enantiomer.
For a non-chiral ensemble, reflection does not change the ensemble, therefore $\alpha^{(K_2K_e)K_J}=0$. 
Interestingly, it can be seen that $\alpha^{(K_2K_e)1}$ can exist in non-chiral molecules for even values of $K_e$.

We now expand the summation over $K_J,N_J$ and insert the explicit density matrix elements for the first photon, giving 
\begin{align}\label{eqn:simplest_general_Z}
\frac{\overline{d\sigma}}{d\mathbf{\hat k}^\mathrm{L}}(E,\tau) \propto  \sum_{ii'}\e^{-\i\omega_{ii'}\tau}\sum_{\substack{K_2K_e}} &-\left[\tfrac{1}{\sqrt{3}}\alpha^{(K_2K_e)0} Z^{(K_2K_e)}_{00}(\mathbf{\hat k}^L)  +\tfrac{1}{\sqrt{6}}\alpha^{(K_2K_e)2} Z^{(K_2K_e)}_{20}(\mathbf{\hat k}^L)\right] \nonumber \\
&-\tfrac{C_1}{\sqrt{2}}\alpha^{(K_2K_e)1} Z^{(K_2K_e)}_{10}(\mathbf{\hat k}^L) \nonumber \\
&+\tfrac{L_1}{2} \alpha^{(K_2K_e)2}\left[ Z^{(K_2K_e)}_{2-2}(\mathbf{\hat k}^L)+
Z^{(K_2K_e)}_{22}(\mathbf{\hat k}^L)  \right],
\end{align}
We remind the reader that this is still the general form for any polarization and propagation direction of the two photons. In the above equation the terms in the first square bracket do not depend on polarization, they exists for an unpolarized pump pulse (note: for a completely unpolarized pump pulse where there is also no preferred propagation direction e.g. produced by three orthogonal beams, only the $Z^{(K_2K_e)}_{00}$ term survives). The term mulitplied by $C_1$ depends on the sign and degree of circular polarization, while the last term depends on the degree of linear polarization.

We now extend the idea of PXECD as described \cite{beaulieu2016, beaulieu2018}. We associate  $C_1\alpha^{(K_2K_e)1}$ as the fundamental quantity describing PXECD, all terms involving it change sign with a change in helicity of the pump pulse, and it only exists when multiple states are coherently populated.
It does not necessarily change sign with enantiomer, and therefore can also exist in non-chiral molecules.
We will see later, when we consider the polarization of the second photon, that the PECD terms arise from $C_2\alpha^{(1K_e)K_J}$ and do not require coherently populated states with non-co-linear dipoles.

To determine the PAD we need to examine
\begin{align}
Z^{(K_2K_e)}_{K_JN_J}(\mathbf{\hat k}^L)=\sum_{\substack{N_eN_2N_2'}}(-1)^{N_2}(\tilde K_J)^{\tfrac{1}{2}}\ThreeJ{K_2}{K_e}{K_J}{-N_2}{N_e}{N_J}\bar \rho^{\xi\mathrm{L}}_{K_2N_2'}\WignerD{K_2}{N_2'}{N_2}(\mu,\nu,\eta) Y_{K_eN_e}(\mathbf{\hat k}^{\mathrm{L}}), 
\end{align}
where we have written $ \rho^{\xi\mathrm{L}}_{K_2N_2}=\sum_{N_2'}\bar\rho^{\xi\mathrm{L}}_{K_2N_2'}\WignerD{K_2}{N_2'}{N_2}(\mu,\nu,\eta)$ and the Euler angles $(\mu,\nu,\eta)$ define the rotation between the coordinate frame of the first and second photon. 

We now look at the specific case of co-propagating pump and probe pulses where the coordinate frame of the two photons coincide, i.e. $\mu=\nu=\eta=0$.
\subsection{Co-propagating pulses}
We obtain the following, 
\begin{align}
 \frac{\overline{d\sigma}}{d\mathbf{\hat k}^\mathrm{L}}(E,\tau) \propto  \sum_{ii'}\e^{-\i\omega_{ii'}\tau}  \left[\frac{1}{3} \alpha_{ii'}^{(00)0}+\frac{1}{6} \alpha_{ii'}^{(20)2}+\frac{1}{2} C_1 C_2 \alpha_{ii'}^{(10)1}+\frac{1}{2} L_1 L_2 \alpha_{ii',\mathrm{S}}^{(20)2}\right]&S_{00}(\mathbf{\hat k}^{\mathrm{L}}) \nonumber \\
 -\left[C_1\left(\frac{\alpha_{ii'}^{(01)1}}{\sqrt{6}}-\frac{ \alpha_{ii'}^{(21)1}}{\sqrt{30}}\right)-C_2\left(\frac{ \alpha_{ii'}^{(11)0}}{3 \sqrt{2}}-\frac{ \alpha_{ii'}^{(11)2}}{3 \sqrt{2}}\right)\right]&S_{10}(\mathbf{\hat k}^{\mathrm{L}})\nonumber \\
 +\left[\frac{\alpha_{ii'}^{(02)2}}{3 \sqrt{2}}+\frac{\alpha_{ii'}^{(22)0}}{3 \sqrt{10}}-\frac{\alpha_{ii'}^{(22)2}}{3 \sqrt{14}}-\frac{C_1 C_2 \alpha_{ii'}^{(12)1}}{\sqrt{10}}+\frac{L_1 L_2 \alpha_{ii'}^{(22)2}}{\sqrt{14}}\right]&S_{20}(\mathbf{\hat k}^{\mathrm{L}})\nonumber \\
 -\left[\frac{C_1 \alpha_{ii'}^{(23)1}}{2} \sqrt{\frac{3}{35}}-\frac{C_2 \alpha_{ii'}^{(13)2}}{2 \sqrt{7}}\right] &S_{30}(\mathbf{\hat k}^{\mathrm{L}})\nonumber \\
 +\left[\frac{\alpha_{ii'}^{(24)2}}{3 \sqrt{14}}+\frac{L_1 L_2 \alpha_{ii'}^{(24)2}}{6 \sqrt{14}}\right]&S_{40}(\mathbf{\hat k}^{\mathrm{L}})\nonumber \\
 -i\sqrt{2}\left[\frac{C_1 L_2 \alpha_{ii'}^{(22)1}}{ \sqrt{5}} + \frac{C_2 L_1 \alpha_{ii'}^{(12)2}}{\sqrt{3}} \right]&S_{2-2}(\mathbf{\hat k}^{\mathrm{L}})\nonumber \\
 -\sqrt{2}\left[L_1\left( \frac{\alpha_{ii'}^{(02)2}}{\sqrt{3}}+\frac{\alpha_{ii'}^{(22)2}}{\sqrt{21}}\right)+L_2\left(\frac{\alpha_{ii'}^{(22)0}}{\sqrt{15}}+\frac{ \alpha_{ii'}^{(22)2}}{\sqrt{21}}\right) \right]&S_{22}(\mathbf{\hat k}^{\mathrm{L}})\nonumber \\
 -i\sqrt{2}\left[\frac{ L_1 \alpha_{ii'}^{(23)2}}{2} \sqrt{\frac{5}{21}}-\frac{L_2 \alpha_{ii'}^{(23)2}}{2} \sqrt{\frac{5}{21}}  \right]&S_{3-2}(\mathbf{\hat k}^{\mathrm{L}})\nonumber \\
  -\sqrt{2}\left[C_2 L_1 \alpha_{ii'}^{(13)2}\sqrt{\frac{5}{42}} -\frac{C_1 L_2 \alpha_{ii'}^{(23)1}}{ \sqrt{14}} \right]&S_{32}(\mathbf{\hat k}^{\mathrm{L}})\nonumber \\
 -\sqrt{2}\left[\frac{L_1 \alpha_{ii'}^{(24)2}}{6} \sqrt{\frac{5}{7}}+\frac{ L_2 \alpha_{ii'}^{(24)2}}{6} \sqrt{\frac{5}{7}} \right]&S_{42}(\mathbf{\hat k}^{\mathrm{L}})\nonumber \\
\end{align}
We can group the terms into 5 classes: Not dependent on light polarization or chirality of the molecule, dependent on circular polarization and chirality, dependent on linear polarization and not chirality, dependent on circular polarization but not chirality, and dependent on linear polarization and chirality. The last two categories are particularly interesting, examples are found, respectively, in the coefficients of $S_{2-2}(\mathbf{\hat k}^{\mathrm{L}})$ which require orientation from the pump(/probe) and alignment from the probe(/pump) and quadrupole emission, and $S_{3-2}(\mathbf{\hat k}^{\mathrm{L}})$ which require alignment of both pump and probe, and octupole emission.
We also see that there are two terms that require both pulses to have circular components involving  $\alpha_{ii'}^{(10)1}$ in the isotropic part of the emission (hence seen in the total cross section) and  $\alpha_{ii'}^{(12)1}$ seen in  $S_{20}(\mathbf{\hat k}^{\mathrm{L}})$, these change sign with change of relative sign between the circularly polarized components of pump and probe pulses, exist for non-chiral molecules, and correspond to both pulses inducing net dipoles in the system, which then couple to give either isotropic or quadrupole emission.

It is also interesting to examine the various coefficients to see their dependence on the orientation/alignment state of the component spherical vectors. 
We notice that terms that change sign due to the circular polarisation of the pump(/probe) pulse always correspond to the orientation vector component induced by the pump(/probe) pulse i.e $K_J(/K_2)=1$.
All terms not dependent on circular polarization have the spherical vectors related to pump and probe pulses as either isotropic or aligned.

We see that asymmetry in the photoemission (corresponding to odd order real spherical harmonics) comes from orientation by the first pulse for PXECD and from the second ionizing pulse for PECD.
We observe that $\alpha_{ii}^{(11)0}$, corresponding to the isotropic part of the first pulse, corresponds to standard one photon PECD from the excited state $i$ up to a constant given by bound transition strength, and so we see that ionization where the second pulse is also circular is not exclusively contingent on coherent population of multiple states.

We can also observe that two-photon PECD (i.e. two circular pulses) mixes PXECD terms with PECD terms in the coefficient of $S_{10}(\mathbf{\hat k}^{\mathrm{L}})$.

We now look at the PXECD experimental setup as described in \cite{beaulieu2016}.

\subsection{Circular pump - linear probe}

Setting $L_1=0$ and $C_2=0$ gives the full angular distribution for PXECD as described in \cite{beaulieu2016}.
The following result is obtained.

\begin{align}
 \frac{\overline{d\sigma}}{d\mathbf{\hat k}^\mathrm{L}}(E,\tau) \propto  \sum_{ii'}\e^{-\i\omega_{ii'}\tau}  \left[\frac{1}{3} \alpha_{ii'}^{(00)0}+\frac{1}{6} \alpha_{ii'}^{(20)2}\right]&S_{00}(\mathbf{\hat k}^{\mathrm{L}}) \nonumber \\
 -\left[C_1\left(\frac{\alpha_{ii'}^{(01)1}}{\sqrt{6}}-\frac{ \alpha_{ii'}^{(21)1}}{\sqrt{30}}\right)\right]&S_{10}(\mathbf{\hat k}^{\mathrm{L}})\nonumber \\
 +\left[\frac{\alpha_{ii'}^{(02)2}}{3 \sqrt{2}}+\frac{\alpha_{ii'}^{(22)0}}{3 \sqrt{10}}-\frac{\alpha_{ii'}^{(22)2}}{3 \sqrt{14}}\right]&S_{20}(\mathbf{\hat k}^{\mathrm{L}})\nonumber \\
 -\left[\frac{C_1 \alpha_{ii'}^{(23)1}}{2} \sqrt{\frac{3}{35}}\right] &S_{30}(\mathbf{\hat k}^{\mathrm{L}})\nonumber \\
 +\left[\frac{\alpha_{ii'}^{(24)2}}{3 \sqrt{14}}\right]&S_{40}(\mathbf{\hat k}^{\mathrm{L}})\nonumber \\
 -i\sqrt{2}\left[\frac{C_1 L_2 \alpha_{ii'}^{(22)1}}{ \sqrt{5}} \right]&S_{2-2}(\mathbf{\hat k}^{\mathrm{L}})\nonumber \\
 -\sqrt{2}\left[L_2\left(\frac{\alpha_{ii'}^{(22)0}}{\sqrt{15}}+\frac{ \alpha_{ii'}^{(22)2}}{\sqrt{21}}\right) \right]&S_{22}(\mathbf{\hat k}^{\mathrm{L}})\nonumber \\
 +i\sqrt{2}\left[\frac{1}{2} \sqrt{\frac{5}{21}} L_2 \alpha_{ii'}^{(23)2} \right]&S_{3-2}(\mathbf{\hat k}^{\mathrm{L}})\nonumber \\
  +\sqrt{2}\left[\frac{C_1 L_2 \alpha_{ii'}^{(23)1}}{ \sqrt{14}} \right]&S_{32}(\mathbf{\hat k}^{\mathrm{L}})\nonumber \\
 -\sqrt{2}\left[\frac{1}{6} \sqrt{\frac{5}{7}} L_2 \alpha_{ii'}^{(24)2} \right]&S_{42}(\mathbf{\hat k}^{\mathrm{L}})\nonumber \\
\end{align}
Or explicitly in vector form
\begin{align}
 \frac{\overline{d\sigma}}{d\mathbf{\hat k}^\mathrm{L}}(E,\tau) \propto  \sum_{ii'}\e^{-\i\omega_{ii'}\tau}  \left[\tfrac{-1}{3\sqrt{3}} \alphazero{00}+\tfrac{1}{6\sqrt{5}} \alphatwo{20}\right]&S_{00}(\mathbf{\hat k}^{\mathrm{L}}) \nonumber \\
 -C_1\left[\tfrac{1}{6}\left(\mathbf{D}^{\mathrm{M}\dagger}_{ii',(01)1}(E) -\tfrac{1}{\sqrt{5}}\mathbf{D}^{\mathrm{M}\dagger}_{ii',(21)1}(E)    \right)\cdot (\mathbf{d}^{M}_{i0} \times \mathbf{d}^{M}_{i'0})\right]&S_{10}(\mathbf{\hat k}^{\mathrm{L}})\nonumber \\
 +\tfrac{1}{3}\left[\left(\tfrac{1}{\sqrt{10}}\mathbf{D}^{\mathrm{M}\dagger}_{ii',(02)2}(E) -\tfrac{1}{\sqrt{70}}\mathbf{D}^{\mathrm{M}\dagger}_{ii',(22)2}(E)\right)\cdot \mathbf{A}_{ii',2} - \tfrac{1}{\sqrt{30}}\alphazero{22} \right]&S_{20}(\mathbf{\hat k}^{\mathrm{L}})\nonumber \\
 -C_1\left[\tfrac{1}{2\sqrt{70}} \alphaone{23}\right] &S_{30}(\mathbf{\hat k}^{\mathrm{L}})\nonumber \\
 +\left[\tfrac{1}{3\sqrt{70}}\alphatwo{24}\right]&S_{40}(\mathbf{\hat k}^{\mathrm{L}})\nonumber \\
 -iC_1 L_2 \left[\tfrac{1}{ \sqrt{15}}\alphaone{22} \right]&S_{2-2}(\mathbf{\hat k}^{\mathrm{L}})\nonumber \\
 +L_2\left[\left(\tfrac{\sqrt{2}}{\sqrt{45}}\alphazero{22}-\tfrac{\sqrt{2}}{\sqrt{105}} \alphatwo{22}\right) \right]&S_{22}(\mathbf{\hat k}^{\mathrm{L}})\nonumber \\
 +iL_2\left[\tfrac{1}{\sqrt{42}}\alphatwo{23} \right]&S_{3-2}(\mathbf{\hat k}^{\mathrm{L}})\nonumber \\
  +C_1 L_2\left[\tfrac{1 }{\sqrt{42}}\alphaone{23} \right]&S_{32}(\mathbf{\hat k}^{\mathrm{L}})\nonumber \\
 - L_2\left[\tfrac{\sqrt{2}}{6\sqrt{7}} \alphatwo{24} \right]&S_{42}(\mathbf{\hat k}^{\mathrm{L}})\nonumber \\
\end{align}
Here we can connect back to the results of \cite{beaulieu2016} where the photoelectron current in the $z$-direction was shown to be a triple product in the Cartesian basis by recognising that $S_{10}(\mathbf{\hat k}^{\mathrm{L}}) \propto k_z$ is responsible for the chiral current in the $z$-direction.
$\left(\mathbf{D}^{\mathrm{M}\dagger}_{ii',(01)1}(E) -\tfrac{1}{\sqrt{5}}\mathbf{D}^{\mathrm{M}\dagger}_{ii',(21)1}(E)\right)$ is equivalent, up to a constant, to the Raman type photoionization vector defined in \cite{beaulieu2016}.
The triple product can be transformed from the  spherical basis to the Cartesian basis by using the usual unitary transformation between the two, this preserves the triple product up to a phase $e^{-i\tfrac{\pi}{2}}$ coming from the determinant of the transformation matrix.
The same transformation can, of course, be applied to all other terms, remembering to multiply by the appropriate phase for transformation of pseudovectors.  
$\mathbf{D}^{\mathrm{M}\dagger}_{ii',(01)1}(E)$ depends only on the isotropic part of the probe pulse and hence survives even for a completely unpolarized probe pulse, while $\mathbf{D}^{\mathrm{M}\dagger}_{ii',(21)1}(E)$ does not.

\section{Conclusions}
We presented a general theory of  PXECD for arbitrary polarization of both the pulse that prepares the molecule in a superposition of excited states, and the ionizing pulse.
A conventional way of analysing angular and energy resolved photoelectron distributions is to perform expansion into the basis of spherical harmonics and analyse the coefficients of this expansion, which are generally referred to as asymmetry parameters.
The theory was developed in a way that clearly and simply separates chiral and non-chiral contributions to the time dependent photoelectron angular distribution for all relevant asymmetry parameters.
PXECD was shown to originate from orientation imposed by the first pulse by inducing a net dipole in the ensemble that oscillates with angular frequency $\omega_{ii'}$ as discussed in \cite{beaulieu2016,beaulieu2018}. The induced chiral dipole underlies the PXCD (photoexcitation circular dichroism) phenomenon introduced in \cite{beaulieu2016,beaulieu2018}.
This is in contrast to one-photon PECD where chiral asymmetric emission  emerges as a result of the orientation imposed by the ionizing pulse.
In PXECD all asymmetry in the forwards/backwards direction (coefficients of and $S_{10}(\mathbf{\hat k}^{\mathrm{L}})$ and $S_{30}(\mathbf{\hat k}^{\mathrm{L}})$ ) is contingent on both chirality and coherent population of multiple states. In contrast, in two-photon PECD there is a mixing of PXCD terms, that require coherent population of excited states,  with PECD-like terms that do not rely on such coherencies.

We have identified the terms uniquely related to PXECD in chiral molecules. We have also shown that PXECD recorded in polarization frame of the pump pulse contains asymmetry parameters, which are exclusively sensitive to coherence, but not associated with chiral response. These terms always arise for field configurations leading to cylindrical symmetry breaking and inducing extrinsic chirality in the polarization plane. 
  
Thus, PXECD is a background-free probe of coherent bound dynamics providing individual access to its chiral and non-chiral contributions.

\section*{Acknowledgements}
We would like to thank useful communications with Andres Ordonez.
AH acknowledges support from DFG project number HA 8552/2-1

\end{document}